\documentclass[conference]{IEEEtran}
\IEEEoverridecommandlockouts
\usepackage{cite}
\usepackage{amsmath,amssymb,amsfonts}
\usepackage{algorithmic}
\usepackage{graphicx}
\usepackage{textcomp}
\usepackage{xcolor}
\usepackage{soul}

\usepackage{ragged2e}
\usepackage{xcolor}
\usepackage{units}
\usepackage{boldline, soul}
\usepackage{multirow}
\usepackage{amsmath}
\usepackage{caption}
\usepackage{pgfplots}
\usepackage{tabularx}
\usepackage{multirow}
\usepackage{hhline}
\usepackage{balance}
\usepackage{booktabs}
\usepackage{balance}
\usepackage{soul}
\usepackage{amssymb}
\usepackage{newtxtext} 

\def\BibTeX{{\rm B\kern-.05em{\sc i\kern-.025em b}\kern-.08em
    T\kern-.1667em\lower.7ex\hbox{E}\kern-.125emX}}

\usepackage{makecell}
\setcellgapes{3pt}
\def\BibTeX{{\rm B\kern-.05em{\sc i\kern-.025em b}\kern-.08em
    T\kern-.1667em\lower.7ex\hbox{E}\kern-.125emX}}

\begin{document}

\title{Merge Mode for Template-based Intra Mode Derivation (TIMD) in ECM}

\author{
    \IEEEauthorblockN{Mohsen Abdoli, Ramin G. Youvalari, Frank Plowman, Alexandre Tissier}
    \IEEEauthorblockA{
        Xiaomi Technology\\
        801 Avenue des Champs Blancs, 35510 Cesson-S\'{e}vign\'{e}, France\\
        Email: mabdoli@xiaomi.com
    }    
}
% \author{Anonymous ICME submission}

\maketitle

\begin{abstract}
This paper presents an intra coding tool, named Merge mode for Template-based Intra Mode Derivation (TIMD). TIMD-Merge has been adopted in the 15\textsuperscript{th} version of the Enhanced Compression Model (ECM) software that explores video coding technologies beyond Versatile Video Coding (VVC) standard. This proposed tool operates on top of the regular TIMD mode that applies a template-based search on a causal adjacent template at top and at left of the current block, in order to find the best Intra Prediction Modes (IPMs) that matches the template. The proposed TIMD-Merge in this paper addresses a shortcoming in the regular TIMD method where due to texture discrepancy, the adjacent template information around the block is not reliable. To do so, the proposed TIMD-Merge constructs a list of all TIMD-coded blocks in relatively larger template area than the template of the regular TIMD, which also includes non-adjacent neighboring blocks. This list, called the merge list, is then sorted on the template to give one best set of TIMD modes. The use of TIMD-Merge mode is signalled at the block level and the implementation in the ECM-14.0 demonstrates -0.08\% performance improvement in terms of luma BDR gain, with negligible encoding and decoding runtime increase of 100.6\% and 100.2\%, respectively. 
\end{abstract}

\begin{IEEEkeywords}
VVC, ECM, Intra Coding
\end{IEEEkeywords}

\section{Introduction}
\label{sec:intro}

While the Versatile Video Coding (VVC) standard approaches industrial adoption \cite{bross2021overview, hamidouche2022versatile}, the Joint Video Experts Team (JVET) is actively developing the next-generation video codec to succeed VVC. This effort aims to identify coding tools that can improve VVC’s compression efficiency \cite{li2024ahg7, abdoli2024stats}. These tools are integrated into a reference software model called the Enhanced Compression Model (ECM), which undergoes continuous refinement to achieve optimal configurations \cite{coban2024ecm}. At the time of this writing, JVET is in the \textit{exploration} phase, laying the groundwork for a formal \textit{standardization} phase expected to conclude around 2029.

The ECM software incorporates numerous advanced tools that enhance compression efficiency compared to VVC. In the realm of intra coding, several innovative tools have been pivotal to these improvements. These include Template-based Intra Mode Derivation (TIMD), Decoder-side Intra Mode Derivation (DIMD) \cite{abdoli2020dimd} and its low-complexity variant, Occurrence-Based Intra Coding (OBIC) \cite{obic}, Spatial Geometric Partition Mode (SGPM) \cite{wang2022sgpm}, Template-based Multiple Reference Line (TMRL) \cite{xu2022tmrl}, intra coding with template matching (IntraTmp) \cite{naser2022itmp}, Convolutional Cross-Component Model (CCCM) \cite{cccm1, 10738512}, Gradient-based Linear Model (GLM) \cite{astola2022glm}, and Cross-Component Prediction (CCP) merge mode \cite{zhang2023ccp}. Over the course of 14 versions developed across four years, ECM-14.0 achieves approximately a 17\% bitrate reduction in intra coding compared to VVC, at the expense of an estimated 962\% increase in encoding time and 507\% increase in decoding time. Concerning inter coding, ECM-14.0 achieves approximately a 27\% bitrate reduction compared to VVC, at the expense of an estimated 922\% increase in encoding time and 1054\% increase in decoding time \cite{ahg12}.

Based on the trade-off of the coding efficiency and complexity performance, the TIMD is one of the most efficient intra coding tools in the ECM. Initially, TIMD was proposed during the exploration phase of the post-HEVC standard \cite{timdpcs} which was not eventually adopted in the VVC due to its high complexity. However, in the post-VVC exploration phase, a reduced complexity version of it was proposed and integrated in the Enhanced Compression Model (ECM) software \cite{cao2021timd}.

This paper introduces an enhancement to the TIMD method that addresses one of its limitations. Specifically, it addresses the scenario where an uncorrelated template around the block is unsuitable for the TIMD to extract prediction information. The proposed TIMD-Merge mode leverages the previously TIMD-coded blocks in the causal neighborhood of a block to derive its Intra Prediction Modes (IPMs). This work was presented at the 36\textsuperscript{th} JVET meeting and subsequently adopted into the ECM-15.0 software. The TIMD-Merge mode has demonstrated consistent improvement, achieving an average BDR luma gain of -0.08\% in the All-Intra configuration and -0.05\% in the Random-Access configuration. 

The remainder of this paper is organized as follows. Section \ref{sec:timd} presents a brief introduction of the TIMD as the basis of the TIMD-Merge mode. Section \ref{sec:timdm} provides detail of the proposed method of this paper. Section \ref{sec:result} presents the experimental results of the TIMD-Merge in the ECM and finally, Section \ref{sec:conclusion} concludes the paper.

% ---------------------------------------------------------------------------------
% ---------------------------------------------------------------------------------
% ---------------------------------------------------------------------------------
% ---------------------------------------------------------------------------------
% ---------------------------------------------------------------------------------
\section{Template-based Intra Mode Derivation (TIMD)}
\label{sec:timd}
In summary, the TIMD algorithm leverages the high correlation between the texture information in the causal neighborhood of the current block and its selected intra prediction mode (IPM). In essence, the TIMD method eliminates signaling of several syntax elements required for the bitrate-intensive IPM coding process, delegating IPM derivation to the decoder-side via an analysis process of a texture that is also available at the decoder side. The TIMD framework comprises four main steps: 1) determining a template area, 2) computing the IPMs' template costs, 3) deriving the best IPMs, and 4) performing block prediction through fusion.

\begin{figure}[t]
\centering
\begin{tabular}{c}
    (a) 
    \\
     \includegraphics[scale=0.42]{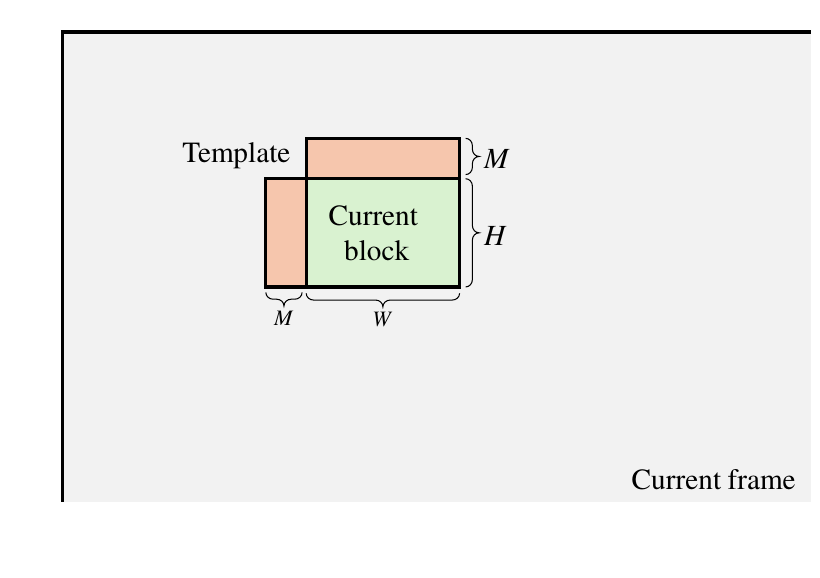}
     \\
     (b)
     \\
     \includegraphics[scale=.55]{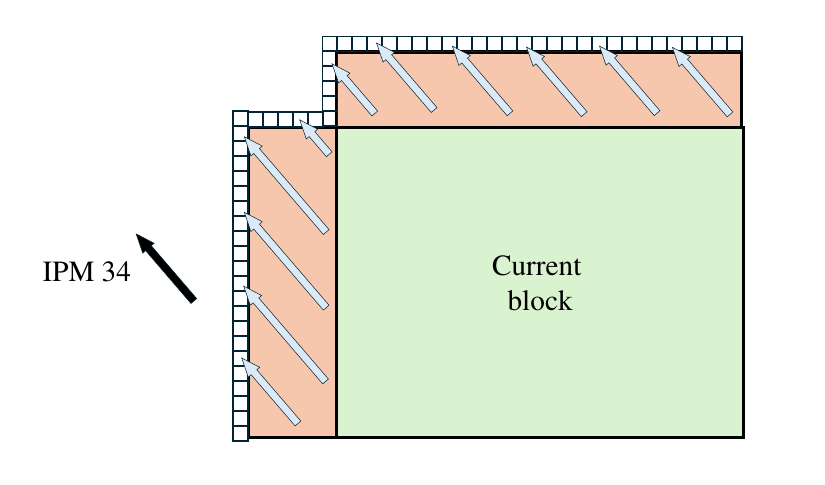}
     
\end{tabular}
\caption{(a) TIMD template with width of $M$ at left and above of the current block of size $W\times H$, (b) Applying template prediction according to the angular IPM 34 and using the template's reference line (white pixels).}
\label{fig:template}
\end{figure}

\noindent
\textit{Determining a template area for IPM prediction}: A template area is selected from a subset of the causal neighborhood of the current block, which is also accessible on the decoder side. The TIMD template consists of two sub-templates, one at the left of the block with the sizes of $H\times M$, and the other one at top of the block with the size $W\times M$, where $W$ and $H$ are current block's width and height, respectively, and $M$ is the number of reference lines on the template (e.g. 2 or 4, depending on the block size). This is illustrated in Fig. \ref{fig:template}-(a). Each sub-template is associated with a reference pixel line, enabling the prediction of the template's pixels using a given IPM. Fig. \ref{fig:template}-(b) provides an example where IPM 34 (diagonal) is applied to both sub-templates of TIMD. The predicted signal generated by an IPM \textit{i} on the template is denoted as $pred_T^i$ throughout this paper.

% \begin{figure*}        
%     \centering
%     \includegraphics[width=0.9\textwidth, angle=0]{figs/template-cropped.pdf}
%     \caption{a) TIMD template with width of $W$ at left and above of the current block, b) Applying template prediction according to the angular IPM 34. }
%     \label{fig:template}
% \end{figure*}

\noindent
\textit{IPM template cost computation}: A subset of IPMs is selected for application in the template area. This subset is derived from the Most Probable Modes (MPM) list and may include variants such as modes with or without the wide-angle option, as well as adjusted-angle versions represented by adjacent IPM indexes. Once the IPM subset is determined, the distortion for each IPM in this subset is computed on the template. A metric known as template cost, denoted as $J$, is defined for a given IPM \textit{i} and template area \textit{T}, and is calculated as:

\begin{equation}
\label{eq:templatecost}
    J_i=M(pred_T^i,recon_T),
\end{equation}
Where $recon_T$ represents the signal containing the reconstructed pixels within the template area, and $M$ is an arbitrary pixel-wise distortion metric. In the current TIMD method implemented in ECM, Sum of Absolute Transform Difference (SATD) is used.

\noindent
\textit{Best IPM derivation}: The next step in the derivation process involves identifying the three best IPMs: two angular IPMs and one non-angular IPM (DC or Planar) that minimize Eq. \ref{eq:templatecost}. These modes are referred to as primary, secondary, and non-angular IPMs of TIMD. For the remainder of this paper, the set of these three modes for a TIMD block is denoted as \textit{D}. 

\noindent
\textit{Block prediction by fusion of derived IPMs}: TIMD performs a fusion of its derived IPMs. Once the three modes are identified, their template costs are used to calculate their fusion weights as follows:

\begin{equation}
    \label{eq:weight1}
    \omega_i=\frac{Sum(J)-J_i}{2\times Sum(J)}, \text{\hspace{10px}} i \in D,
\end{equation}
where $Sum(J) = \sum_i J_i$ for $i \in D$. In other words, Eq. \ref{eq:weight1} allocates higher fusion weights for IPMs with smaller costs $J_i$, by comparing them to other IPMs through the value of $Sum(J)$ that stores sum of all costs.

Finally, pixels within a block are predicted by generating a prediction signal, denoted as $pred_B$. This signal is obtained by computing three prediction signals for the pixel positions within the block and then applying a linear combination of these signals using their corresponding fusion weights. Eq. \ref{eq:fusion} denotes this process:

\begin{equation}
    \label{eq:fusion}
    pred_B=\Sigma_i \omega_ipred_B^i, \text{\hspace{10px}} i \in D,
\end{equation}
where $pred_B^i$ denotes the prediction signal for each individual IPM in \textit{D}, computed on the pixels within the \textit{current block}. Note that $pred_B^i$ is distinct from $pred_T^i$, which is computed on the pixels within the \textit{template area}, as denoted in Eq. \ref{eq:templatecost}.

\noindent
\textit{Template discrepancy problem in TIMD}:
Although using the causal neighborhood as template to derive the IPM helps TIMD achieve significant bitrate savings, the template information can be misleading in various scenarios. For example, texture orientation mismatches, repetitive patterns in the scene, and structural noise can lead to issues for TIMD. Under such conditions, the performance of TIMD is negatively impacted due to discrepancies between the template's content and the block's content. This paper attempts to address this issue by introducing the TIMD-Merge mode. In the rest of this paper, the TIMD method described above is referred to as “\textit{regular TIMD}”, to distinguish from its proposed sub-mode TIMD-Merge method.

% ---------------------------------------------------------------------------------
% ---------------------------------------------------------------------------------
% ---------------------------------------------------------------------------------
% ---------------------------------------------------------------------------------
% ---------------------------------------------------------------------------------
\section{Merge mode for TIMD}
\label{sec:timdm}
The TIMD-Merge algorithm reduces the reliance on uncorrelated template texture and allows for the exploitation of potential texture correlations with additional blocks in the causal neighborhood. These correlated blocks are typically not accessible to the regular TIMD due to its small template area. The proposed TIMD-Merge mode consists of three main steps: 1) constructing the merge list, 2) reordering candidates based on the template, and 3) inheriting prediction information from the best candidate.

\noindent
\textit{Merge list construction}: In the first step, a merge list $L$ is constructed based on the blocks in the neighborhood, consisting of adjacent and non-adjacent neighboring blocks. In both cases, a set of predefined positions is used to identify unique neighboring blocks that are eligible to serve as merge candidates in the TIMD-Merge list $L$. In the proposed method, an eligible candidate is a block that is coded either with regular TIMD or TIMD-Merge mode.
% @NOTE(frank) This previously read "As for adjacent blocks, left and above positions of the current block are scanned with a step size of 4 pixels."

\begin{figure}[t]
\centering
\begin{tabular}{c}
     (a)
     \\
     \includegraphics[scale=0.32]{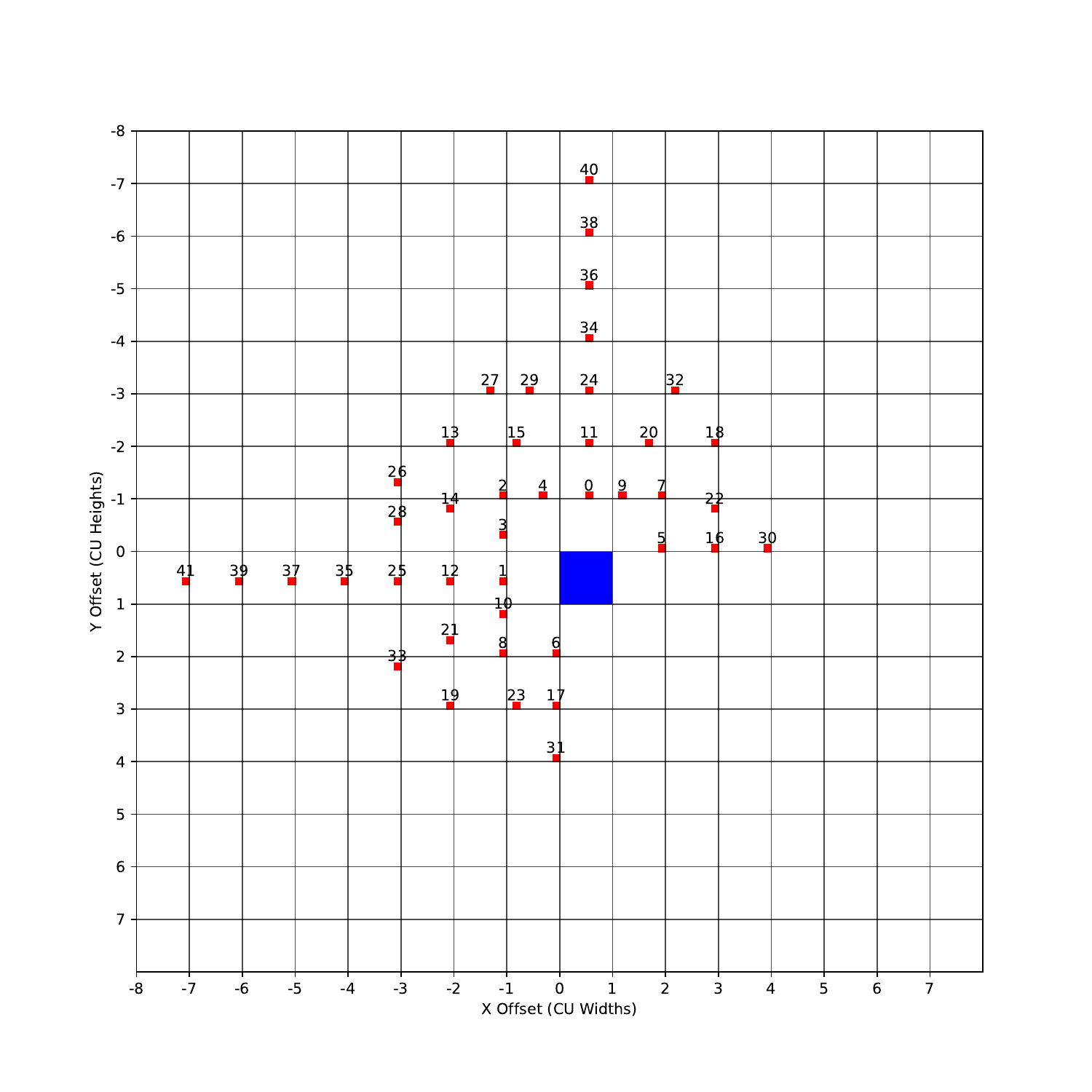}
     \\
     (b)
     \\
     \includegraphics[scale=.55]{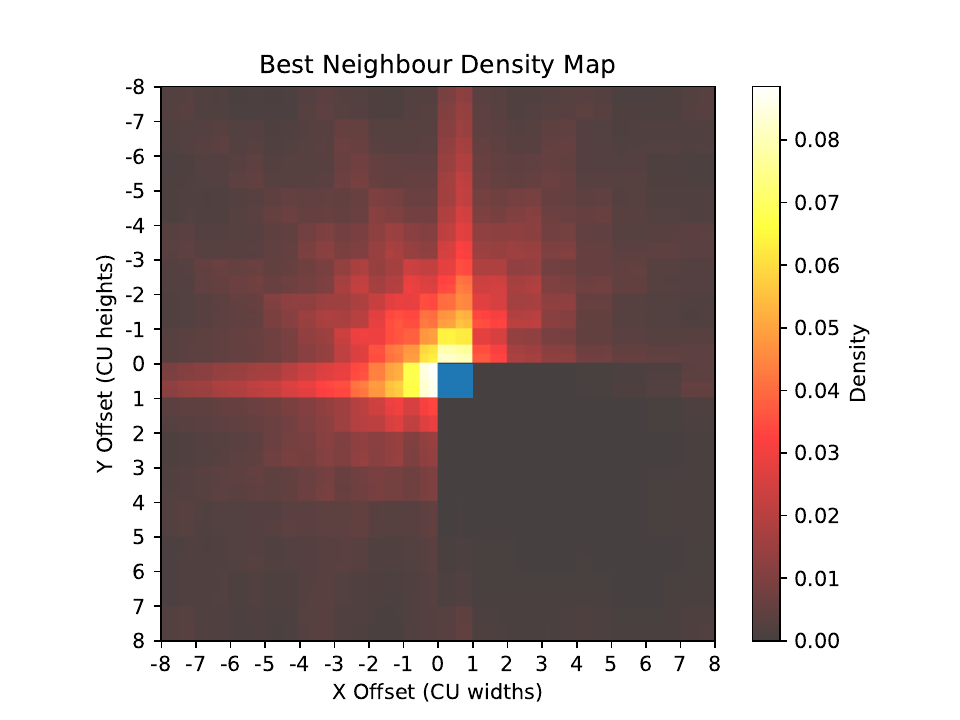}
\end{tabular}
\caption{(a) Map of the 41 candidate non-adjacent neighbor positions. (b) Proportion of TIMD-merge-coded CUs which use prediction information from a CU covering the position at the relevant normalized offset, when using the non-adjacent neighbor map.}
\label{fig:map}
\end{figure}

The adjacent neighboring blocks comprise every previously-decoded CU which shares an edge with the current block. These are found by checking every 4\textsuperscript{th} sample on the left and top borders of the current CU, where 4 is the minimum CU size in ECM.
The non-adjacent neighbors are selected based on a neighborhood map, as depicted in Fig. \ref{fig:map}-(a). In this figure, it is important to note that the grid in Fig. \ref{fig:map}-(a) is normalized according to the size and aspect ratio of the current block at its center. In other words, the width and height of each cell in the grid is equal to the width and height of the current block. For blocks of different sizes and/or aspect ratios, the grid and neighbor positions are adjusted accordingly.

% \begin{figure}        
%     \centering
%     \includegraphics[width=0.45\textwidth, angle=0]{figs/map.pdf}
%     \caption{Map of the 41 candidate non-adjacent neighbor positions. %There are also $(w + h)/4 + 7$, candidate adjacent neighbor positions on the CUs neighboring the current block, where $w$ and $h$ denote the width and height of the current block.
%     }
%     \label{fig:nbmap}
% \end{figure}

The non-adjacent neighborhood map used for TIMD-Merge has been empirically optimized based on an statistical analysis of correlation between best IPM of current block and its neighbors. The heat-map in Fig. \ref{fig:map}-(b) shows the proportion (density) of TIMD-merge-coded blocks which inherit prediction information from a block containing the (normalized) offset position, when using a different non-adjacent neighbor map. This map has more candidates than TIMD-merge's, and for the purposes of this analysis can be considered a covering of the neighborhood. For example, Fig. \ref{fig:map}-(b) shows that positions in the top-right and bottom-left quadrants, as well as far from the current block in the top-left quadrant, are unlikely to have prediction information relevant to the current block and therefore need not be checked. TIMD-merge's non-adjacent neighbor map shown in Fig. 2 was designed to have a density of candidate positions similar to the probability that neighbors at the respective offset will be useful.
% \begin{figure}        
%     \centering
%     \includegraphics[width=0.45\textwidth, angle=0]{figs/heatmap.pdf}
%     \caption{Proportion of TIMD-merge-coded CUs which use prediction information from a CU covering the position at the relevant normalized offset, when using the dense non-adjacent neighbor map.}
%     \label{fig:heatmap}
% \end{figure}

\noindent
\textit{Candidate reordering on template}: Even though the template around current block is assumed to be less reliable in case of TIMD-Merge than the regular TIMD, there is still relevant information in the template that can be exploited. In particular, when the merge list is constructed from all eligible adjacent and non-adjacent neighbor blocks, the template cost is used in a limited way (only one reference line) to reorder the candidates. To this end, first template costs of individual IPMs in all candidates of the merge list are computed, using Eq. \ref{eq:templatecost}. For each candidate \textit{c} in the merge list, with a set of derived IPMs denoted as $D_c$ (containing its primary, secondary and non-angular IPMs), the relative weights, $\omega_i$, of derived IPMs ($i \in D_c$) are computed using Eq. \ref{eq:weight1}. Then, a metric named the candidate cost is computed as follows:
\begin{equation}
    \label{eq:candcost}
    K_c=\Sigma_iJ_i.\omega_i, \text{\hspace{10px}} i\in D_c
\end{equation}

The candidate reordering is applied by sorting the candidates in the merge list based on their computed candidate cost. Finally, the best candidate that minimizes Eq. \ref{eq:candcost}, and is selected for coding current block with TIMD-merge:

\begin{equation}
\label{eq:opt}
C^* = \arg\min_{C \in L} K_c.
\end{equation}

\noindent
\textit{Inherit Prediction information}: Once the best candidate $C^*$ is identified, prediction information are inherited for the current block. This inherited information are namely the set of TIMD modes $D_{c^*}$ (consisting its primary, secondary and non-angular IPM) and their fusion weights. 

Moreover, the horizontal and vertical transform types of $C^*$ are also inherited for current block. The reasoning behind inheriting the transform types is that if the content of the current block is similar to that of the candidate -- which is why current block is trying the merge mode -- then it is likely there are also residual characteristic similarities between the two blocks so inheriting the transform type could further reduce the bitrate needed for its explicit signaling.

After the prediction inherence step, the rest of the prediction process by the proposed TIMD-Merge method is the same as in the regular TIMD. This consists of computing individual prediction signals ($pred_B^i, i \in D_{c^*}$) and finally block prediction by fusion (Eq. \ref{eq:fusion}). 

\begin{figure}        
    \centering
    \includegraphics[width=0.45\textwidth, angle=0]{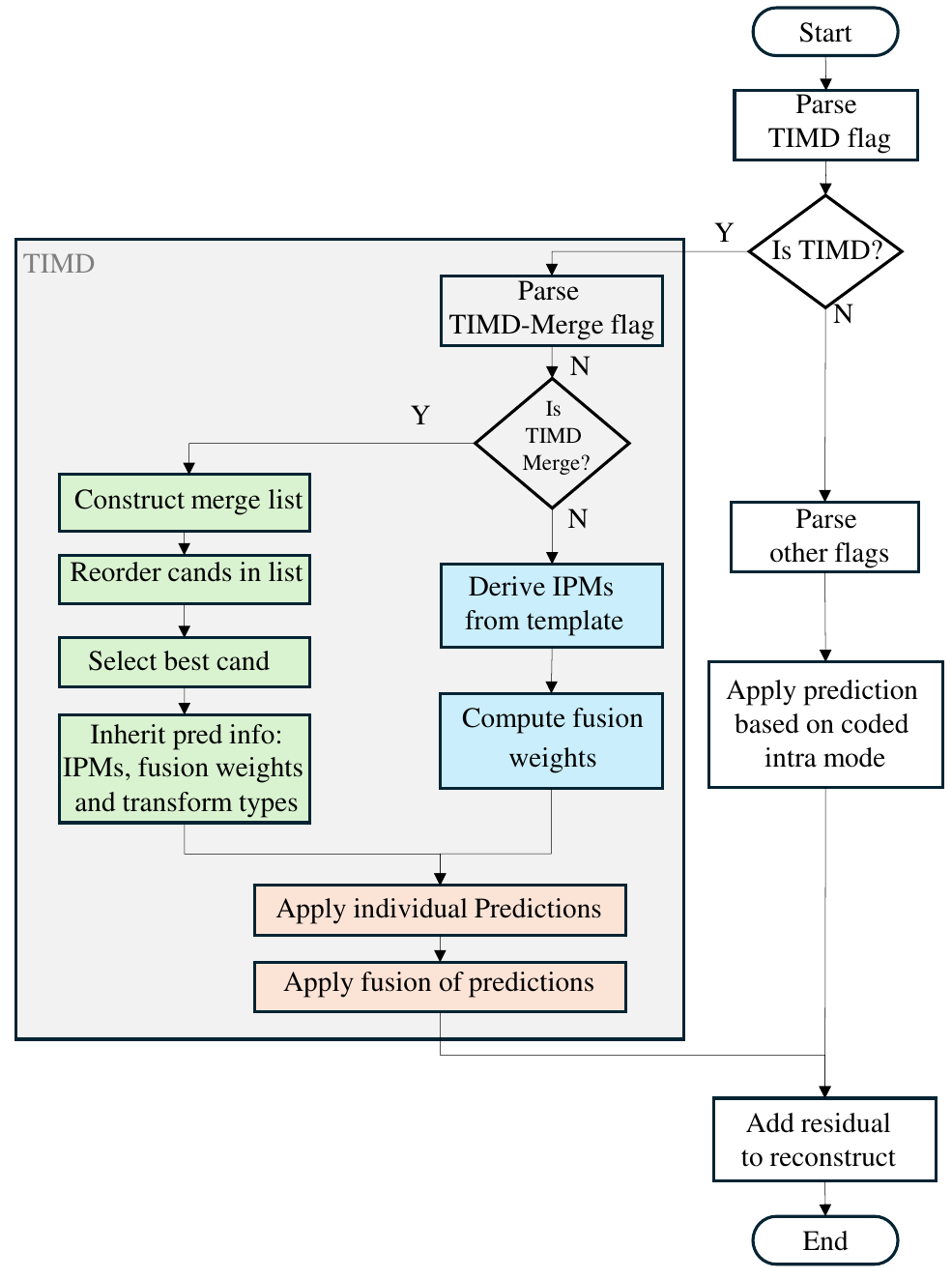}
    \caption{Diagram of the ECM decoder with proposed modification of TIMD. Green: Aspects specific to regular TIMD. Blue: Aspects specific to TIMD-Merge. Orange: Shared aspects.}
    \label{fig:diagram}
\end{figure}

Fig. \ref{fig:diagram} shows the diagram of the ECM decoder after integration of the proposed TIMD-Merge algorithm. As can be seen, TIMD-Merge is placed as a sub-mode of the regular TIMD and is distinguished by a dedicated block-level flag. Depending on the parsed value of this flag, the decoder determines whether TIMD prediction information are derived from the template (blue) or they are inherited from the best candidate of the merge list (green). Once the prediction information are determined, both methods share the same process to compute the individual predictions and apply a fusion on the individual predictions (orange). At the encoder side, in order to control the encoding time, the proposed TIMD-Merge mode is first compared to the regular TIMD mode in terms of Sum of Absolute Transform Difference (SATD). In this comparison, if the SATD cost of the TIMD-Merge mode is significantly worse than the regular TIMD mode, it is not passed to the full RD cost computation.

% ---------------------------------------------------------------------------------
% ---------------------------------------------------------------------------------
% ---------------------------------------------------------------------------------
% ---------------------------------------------------------------------------------
% ---------------------------------------------------------------------------------
\section{Experiment result}
\label{sec:result}
To assess the performance of the proposed TIMD-Merge method, it has been integrated into the 14\textsuperscript{th} version of the Enhanced Compression Model (ECM). As discussed in the previous section, this integration required changes in the signaling of the intra coding flag. Specifically, when the TIMD flag is enabled, a second block-level flag is introduced to indicate whether the regular TIMD or the merge mode of TIMD is used. The proposed TIMD-Merge method described in this paper was presented at the 33\textsuperscript{rd} JVET meeting and subsequently adopted into ECM-15.0 software at the 36\textsuperscript{th} JVET meeting \cite{abdoli2024timdmerge}. All tests were carried out using the encoder configuration and sequences defined in the Common Test Conditions (CTC) recommended by JVET \cite{Karczewicz2023ctc}. Results are provided in the All-Intra (AI) as well as Random-Access (RA) configurations. 

\begin{figure*}[t]
\begin{tabular}{ccc}
    \small{(c) BQSquare ($416\times 240$)} & \small{(a) BQMall ($832\times 480$)} & \small{(b) ParkRunning ($3840\times 2160$)}
    \\
     \includegraphics[scale=0.52]{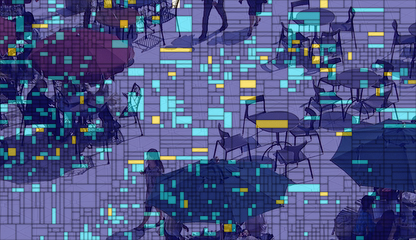}
     &
     \includegraphics[scale=0.26]{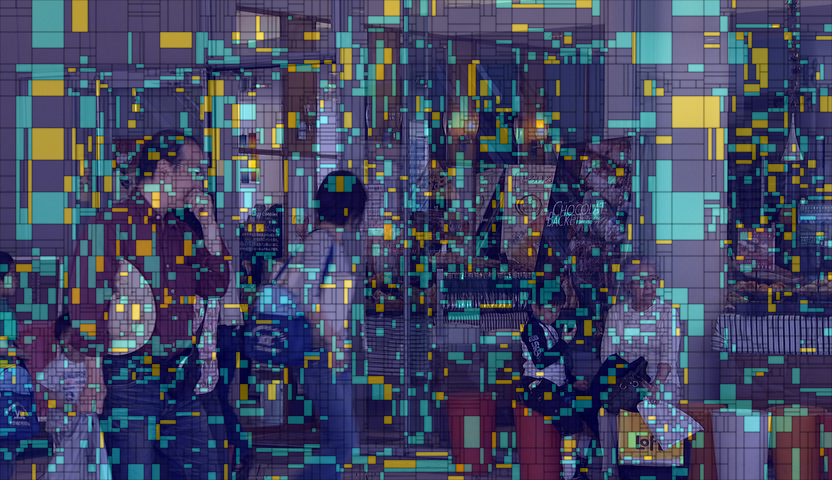}
     &
     \includegraphics[scale=0.265]{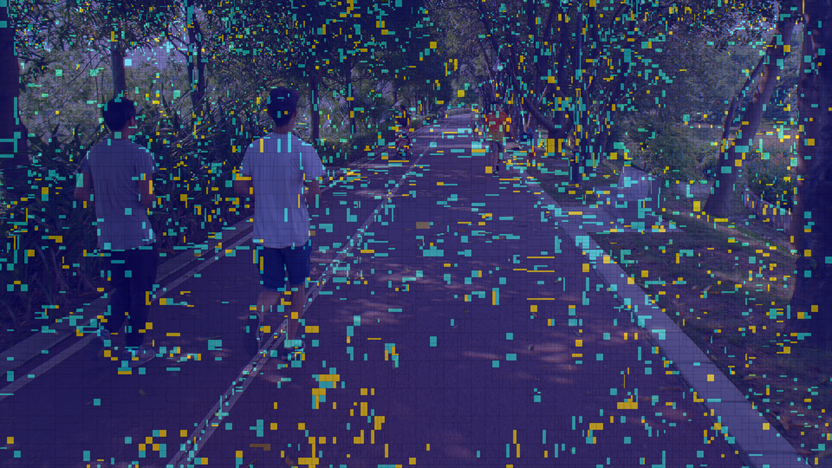}
\end{tabular}
\caption{Visual demonstration of blocks selected by the regular TIMD (cyan), the proposed TIMD-Merge (yellow) and other intra tools (purple).}
\label{fig:selection}
\end{figure*}

\begin{table*}[h!]
	\centering
	\caption{BD-Rate (\%) performance of TIMD-Merge tool over ECM-14.0}
	\label{tab:bdr}
        %\begin{tabular}{c c c c|}
         %&  All Intra &  & \\
         %\hline
	\begin{tabular}{c|c|c c c||c c c}
            \hline
            \multicolumn{8}{c}{\hspace{3.5cm} All Intra \hspace{2cm} Random Access}\\
            \hline
		Class & Sequence & Y & Cb & Cr & Y & Cb & Cr\\
		\hline
		A1 &Tango   &  -0.09\%	&-0.28\%	&0.01\%& -0.01\% &	-0.51\% &	-0.55\% \\
		&FoodMarket      &  -0.09\%	&-0.01\%	&-0.07\%  &-0.03\% &	-0.26\% &	0.05\%  \\
		&CampfireParty          & -0.06\%	&0.06\%	&-0.15\% & -0.05\% &	0.01\% &	-0.12\%  \\
  \hline
		A2 &CatRobot     &  -0.11\%	&-0.10\%	&0.00\%  & -0.02\% &	-0.11\% &	-0.01\%  \\
		&DaylightRoad        & -0.07\%	&-0.08\%	&-0.17\%  & -0.11\% &	-0.30\% &	0.11\%  \\
		&ParkRunning& -0.06\%	&0.06\%	&-0.04\%  & 0.00\% &	0.00\% &	0.00\% \\
  \hline
		B &MarketPlace       & -0.06\%&	0.19\%&	0.09\%  & -0.01\% &	0.39\% &	-0.18\%  \\
		&RitualDance         &  -0.06\%	&-0.26\%&	-0.01\%  & -0.03\% &	-0.16\% &	0.06\%  \\
		&Cactus        & -0.09\%&	0.05\%&	0.01\%  & -0.04\% &	-0.09\% &	-0.08\% \\
        &BasketballDrive        & -0.13\%&	0.44\%&	0.02\% & -0.03\% &	-0.41\% &	-0.02\%  \\
        &BQTerrace        & -0.09\%&	-0.06\%&	0.47\%  & -0.16\% &	-0.09\% &	-0.47\%  \\
		\hline
		C &BasketballDrill  & -0.11\%&	-0.13\%	&0.00\% & -0.05\% &	0.15\% &	0.17\%  \\
		&BQMall         & -0.10\%&	-0.12\%	&-0.20\% & -0.05\% &	-0.49\% &	-0.33\% \\
		&PartyScene       & -0.07\%&	-0.21\%&	-0.07\% & -0.04\% &	-0.19\% &	0.36\%  \\
		&RaceHorses          & -0.04\%&	-0.07\%	&0.26\% & 0.04\% &	-0.25\% &	0.29\% \\
  \hline
		D &BasketballPass        & -0.09\%&	0.07\%&	0.06\% & 0.02\% &	-0.24\% &	-0.23\%  \\
		&BQSquare      & -0.11\%&	0.28\%&	0.25\% & -0.05\% &	-0.70\% &	-0.84\%  \\
		&BlowingBubbles   & -0.07\%	&0.01\%&	-0.37\%  & -0.04\% &	-0.07\% &	0.00\%  \\
		&RaceHorses  	 & 0.02\%&	-0.21\%&	0.08\% & 0.00\% &	-0.43\% &	-0.43\% \\
  \hline
  		E &FourPeople        & -0.09\%&	0.01\%&	0.04\% & -& -& -\\
		&Johnny      & -0.09\%&	0.52\%&	0.85\%  & -& -&-\\
		&KristenAndSara   & -0.07\%&	0.23\%	&-0.07\%   & -& -& -\\
\hline
        F &BasketballDrillText        & 0.06\%&	0.03\%&	-0.07\% & - &	- &	-  \\
		&ArenaOfValor      & 0.02\%&	-0.15\%&	0.01\% & - &	- &	-  \\
		&SlideEditing   & -0.04\%	&0.10\%&	0.11\%  & - &	- &	-  \\
		&SlideShow  	 & -0.23\%&	-0.56\%&	-0.32\% & - &	- &	- \\
\hline
        TGM &FlyingGraphincs        & -0.23\%&	-0.12\%&	-0.08\% & - &	-  &	- \\
		&Desktop      & -0.12\%&	-0.17\%&	-0.01 & - &	- &	-  \\
		&Console   & -0.03\%	&-0.11\%&	0.08\%  & - &	- &	-  \\
		&ChineseEditing  	 & -0.08\%&	0.06\%&	-0.12\% & - &	- &	- \\
\hline
        \multicolumn{2}{c}{{Average class A1}} & {-0.08\%} & {-0.07\%} &	{-0.07\%} & {-0.03\%}& {-0.25\%} &{-0.21\%}\\ 
		\hline
		\multicolumn{2}{c}{{Average class A2}}  & {-0.08\%} & {-0.04\%} & {-0.07\%} &	{-0.04\%} & {-0.14\%}& {0.03\%}\\ 
		\hline
  		\multicolumn{2}{c}{{Average class B}}  & {-0.09\%} & {0.07\%} & {0.11\%} &	{-0.05\%} & {-0.07\%}& {-0.14\%}\\ 
		\hline
  		\multicolumn{2}{c}{{Average class C}}  & {-0.08\%} & {-0.13\%} & {0.00\%} &	{-0.02\%} & {-0.19\%}& {0.12\%}\\ 
		\hline
  		\multicolumn{2}{c}{{Average class D}}  & {-0.06\%} & {0.03\%} & {0.00\%} &	{-0.02\%} & {-0.36\%}& {-0.38\%}\\ 
		\hline
  		\multicolumn{2}{c}{{Average class E}}  & {-0.08\%} & {0.25\%} & {0.27\%} &	{-} & {-}& {-}\\
        \hline
        \multicolumn{2}{c}{{Average class F}}  & {-0.05\%} & {-0.19\%} & {-0.07\%} &	{-} & {-}& {-}\\
        \hline
        \multicolumn{2}{c}{{Average class TGM}}  & {-0.12\%} & {-0.09\%} & {-0.03\%} &	{-} & {-}& {-}\\
		\hline
		\multicolumn{2}{c}{\textbf{Overall}} & \textbf{-0.08\%} & \textbf{0.01\%} &  \textbf{0.05\%} & \textbf{-0.05\%}& \textbf{-0.07\%}& \textbf{-0.11\%}\\
            \hline
            \multicolumn{2}{c}{Average Enc. Time} & \multicolumn{3}{c||}{100.6\%} & \multicolumn{3}{c}{99.9\%} \\
            \multicolumn{2}{c}{Average Dec. Time} & \multicolumn{3}{c||}{100.0\%} & \multicolumn{3}{c}{100.1\%} \\
		\hline
	\end{tabular}
 %\end{tabular}
\end{table*}

% \begin{figure}[t]
% \begin{tabular}{c}
%     \small{(a) BQMall }
%     \\
%      \includegraphics[scale=0.38]{figs/overlays/BQMallQ22_frm0.png}
%      \\ 
%      \small{(b) ParkRunning}
%      \\
%      \includegraphics[scale=0.38]{figs/overlays/parkrunningq37_frm0_resizedC.png}
%      \\
%      \small{(c) BQSquare }
%      \\
%      \includegraphics[scale=0.76]{figs/overlays/bqsquare_q22_frm3.png}
% \end{tabular}
% \caption{Visual demonstration of blocks selected by the regular TIMD (cyan), the proposed TIMD-Merge (yellow) and other intra tools (purple).}
% \label{fig:selection}
% \end{figure}

\begin{figure*}[t]
    \centering
    \includegraphics[width=0.95\textwidth, angle=0]{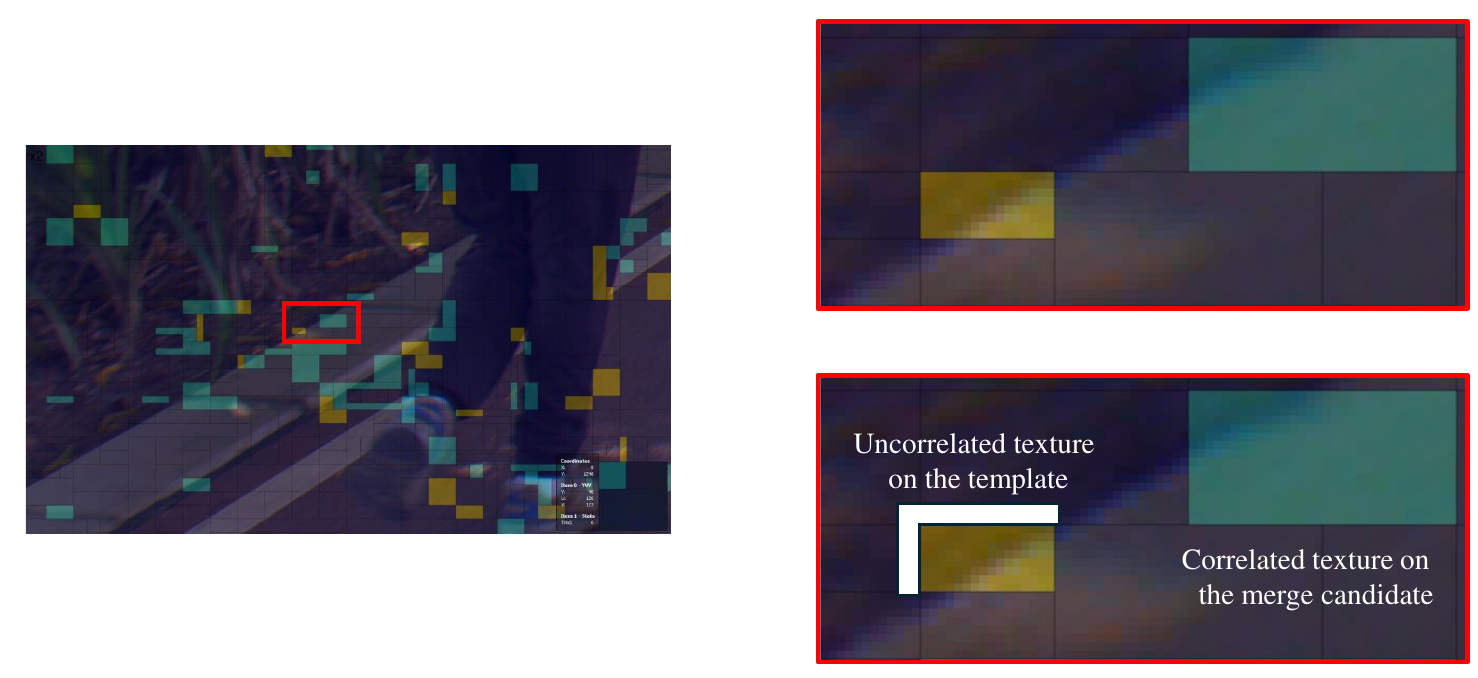}
    \caption{Example of a TIMD-Merge block (yellow) with uncorrelated template (white), inheriting prediction information from its correlated non-adjacent neighbor (cyan) that is coded with the regular TIMD.}
    \label{fig:visblock}
\end{figure*}

Table \ref{tab:bdr} presents the Bj\o{}ntegaard-Delta Rate (BDR) \cite{bjontegaard2001calculation} as well as the runtime measures for both the encoder and decoder sides. In the AI configuration, the overall luma BDR gain is -0.08\%, with encoding and decoding time increases of 100.6\% and 100.2\%, respectively. As shown, the achieved gain is consistent across all CTC classes, regardless of resolution or content characteristics. Additionally, since both the regular TIMD and the proposed TIMD-Merge methods are applied exclusively to the luma component, no significant benefit is observed for the chroma components. In an ablation study was conducted to assess the impact of adjacent and non-adjacent neighbors in the neighborhood map. The experiments indicate that excluding non-adjacent neighbors from the TIMD-Merge process leads to a significantly smaller gain of -0.03\%. In another ablation study, the transform inheritance aspect was deactivated, which resulted in a smaller gain of -0.04\%.

The increase in encoding time is due to testing the proposed TIMD-Merge method as an additional coding mode at the encoder side. However, the design of TIMD-Merge mitigates this by inheriting the transform types from the best merge candidate, so the encoder does not need to perform additional tests to determine these transform types. At the decoder side, the slight increase in decoding time is primarily attributed to the construction of the merge list, specifically the cost computation on the template for each individual IPM.

Fig. \ref{fig:selection} provides a visual representation of how blocks coded using regular TIMD (cyan), the proposed TIMD-Merge (yellow), and other intra coding tools (purple) are spatially distributed. The TIMD-Merge method is consistently chosen across various texture types, block sizes, and shapes.

Fig. \ref{fig:visblock} illustrates an example from the ParkRunning sequence. To reiterate, TIMD-Merge aims to enhance the performance of regular TIMD, particularly when the texture in the immediate template is uncorrelated with the texture of the current block, yet there are correlated textures in the broader neighborhood that are inaccessible to regular TIMD. In this example, the yellow block is coded using TIMD-Merge, where the L-shaped template around it belongs to a different object, making it unsuitable for IPM derivation. However, the cyan block coded with regular TIMD in its neighborhood provides the correlated prediction IPM information for the current block. As a result, the current block can benefit from the advanced coding capabilities of regular TIMD (e.g. fusion), even without having a directly correlated template.

An empirical analysis of the texture characteristics and RD cost of the proposed TIMD-Merge method suggests that there are instances where it may fail to accurately code a block. In particular, the authors observed that when a sudden change occurs in the neighboring texture (e.g., due to a transition between two objects), TIMD-Merge may fail if this change happens earlier within its immediate template. In other words, if the immediate template belongs to the same object as the current block, TIMD-Merge attempts to retrieve relevant IPM information from the previous object, leading to the derivation of poorly correlated information.    

% ---------------------------------------------------------------------------------
% ---------------------------------------------------------------------------------
% ---------------------------------------------------------------------------------
% ---------------------------------------------------------------------------------
% ---------------------------------------------------------------------------------
\section{Conclusion}
\label{sec:conclusion}
The TIMD-Merge tool presented in this paper offers an enhancement to the regular TIMD method by addressing its limitations related to texture discrepancies in template-based intra prediction. By bypassing the uncorrelated template area and incorporating textures in adjacent and non-adjacent neighboring blocks, TIMD-Merge improves the IPM derivation performance of the TIMD. The implementation of TIMD-Merge in the ECM-14.0 software has shown a performance improvement, achieving a -0.08\% BDR gain while maintaining manageable increases in encoding and decoding runtimes. The work presented in this paper has adopted in the 15\textsuperscript{th} version of the ECM software for exploration of the post-VVC standard.

% ---------------------------------------------------------------------------------
% ---------------------------------------------------------------------------------
% ---------------------------------------------------------------------------------
% ---------------------------------------------------------------------------------
% ---------------------------------------------------------------------------------
\balance
\bibliographystyle{IEEEbib}
\bibliography{icme2025references}

\end{document}